%% file: timeLag.tex
\documentclass[
 aps, pra,
 amsmath,amssymb,
 11pt,
 final,
tightenlines,
 twoside,
 twocolumn,
 nofloats,
nofootinbib,
 superscriptaddress,
showkeys,
showkeywords,
 ]
{revtex4-2}
\usepackage[T2A]{fontenc}
\usepackage{natbib}
\usepackage[utf8x]{inputenc}
\usepackage[english]{babel}
\usepackage{graphicx}% Include figure files
\usepackage{dcolumn}% Align table columns on decimal point
\usepackage{bm}% bold math

\input{maik.rty}

\setcitestyle{authoryear,round}
\input{sao_cmd_author.tex}

\begin{document}

\selectlanguage{english}

\title{Time Lag between Accretion and Wind Events in the T Tauri Star RY Tau}

\author{\firstname{E.~V.}~\surname{Babina}}
 \email{helenka_truth@mail.ru}
 \affiliation{Crimean Astrophysical Observatory, Republic of Crimea, 298409, Russia}

\author{\firstname{P.~P.}~\surname{Petrov}}
\affiliation{Crimean Astrophysical Observatory, Republic of Crimea, 298409, Russia}

\author{\firstname{K.~N.}~\surname{Grankin}}
\affiliation{Crimean Astrophysical Observatory, Republic of Crimea, 298409, Russia}

\author{\firstname{S.~A.}~\surname{Artemenko}}
\affiliation{Crimean Astrophysical Observatory, Republic of Crimea, 298409, Russia}

\begin{abstract}

The results of spectroscopic and photometric monitoring of the classical T Tauri star RY Tau are presented. The observation series span 220~nights from 2013 to 2024. During the observation period, the star's brightness varied within the range of V=9-11$^{m}$. The rotation axis of the “star + accretion disk”~system is tilted at a large angle, so the line of sight intersects the wind region and accreting flows in the star's magnetosphere. Variability in the short-wavelength wing of the H$\alpha$ emission line and the profile of the D\,NaI resonance doublet are analyzed. It is shown that the wind and accretion flows vary on a time scale of approximately 20~days. When the predominant flow direction changes, a time lag is observed: initially, accretion increases, and after two days, absorption in the line-of-sight wind decreases. It is concluded that the spectral line profiles are formed in the magnetospheric accretion flows and the \textit{conical wind} originating from the boundary of the star's magnetosphere. The time lag is determined by the tilt of the magnetic dipole and the opening angle of the conical wind. It is assumed that RY Tau operates in an unstable propeller mode, and fluctuations in the accretion and wind flows are caused by density waves in the accretion disk.


\end{abstract}

\maketitle

\section{Introduction}

%Классические переменные типа Т~Тельца (CTTS) -- молодые звезды малых масс ($\leq~2M_\odot$) с аккреционными дисками  -- показывают как признаки  продолжающейся аккреции, так и признаки интенсивного ветра. Модель магнтосферной аккреции \citep{Cam1990, Koe1991, Shu1994, HarHew1994} %(Camenzind et al. 1990; Koenigl 1991; Shu et al. 1994;  Hartmann et al. 1994) 
%~наиболее успешно объясняет спектральные  особенности этих звезд, включая профили эмиссионных линий, свидетельствующие  о потоках газа к звезде (аккреция) и от звезды (ветер)  на луче зрения \citep{AlBas2000}. %(Alencar Basri 2000).  
%~Согласно этой модели, дисковая аккреция на молодую звезду останавливается магнитным полем  звезды на расстоянии, которое определяется темпом аккреции и напряженностью дипольной  составляющей магнитного поля звезды. С этого расстояния   происходит свободное падение газа на поверхность звезды вдоль силовых линий  магнитосферы. В месте падения на поверхности звезды возникает ударная волна с температурой  \sim10$^6$~К.  Большая часть УФ излучения ударной волны сосредоточена  в эмиссионной линии  Ly$\alpha$ \citep{ArGrF2023}.% (Arulanantham еt al, 2023).  
%~Рентгеновское и УФ излучение ударной волны ионизует газ в окрестности звезды  и нагревает  лежащий ниже участок фотосферы -- горячее пятно. Излучением горячего пятна объясняется наблюдаемый УФ избыток в  распределении энергии в спектре CTTS  и эффект вуалирования фотосферных линий \citep{DoLam2012}.% (Додин и Ламзин 2012). 
%~Темп аккреции CTTS составляет  10$^{-8}$--10$^{-9}$M_$\odot/yr$ \citep{BouGrAl2007}. %(Bouvier et al. 2007).

Classical T~Tauri variables (CTTS) are young, low-mass stars ($\leq~2M_\odot$) with accretion disks that show both signs of ongoing accretion and signs of an intense wind. The magnetospheric accretion model \citep{Cam1990, Koe1991, Shu1994, HarHew1994}%(Camenzind et al. 1990; Koenigl 1991; Shu et al. 1994; Hartmann et al. 1994) 
~most successfully explains the spectral features of these stars, including the emission line profiles indicating line-of-sight gas flows toward the star (accretion) and away from the star (wind) \citep{AlBas2000}.%(Alencar & Basri 2000). 
~According to this model, disk accretion onto the young star is stopped by the star's magnetic field at a distance determined by the accretion rate and the strength of the dipole component of the stellar magnetic field. From this distance, free fall of gas occurs onto the star's surface along the magnetospheric field lines. At the impact site, a shock wave with a temperature of $\sim10^6$~К is generated on the star's surface. A significant portion of the shock wave's UV radiation is concentrated in the Ly$\alpha$ emission line \citep{ArGrF2023}.%(Arulanantham et al., 2023). 
~The X-ray and UV radiation from the shock wave ionizes gas in the star's vicinity and heats the underlying region of the photosphere-the hot spot. The hot spot radiation explains the observed UV excess in the energy distribution in the CTTS spectrum and the veiling effect of photospheric lines \citep{DoLam2012}.%(Dodin and Lamzin 2012). 
~The CTTS accretion rate is 10$^{-8}$--10$^{-9}$~M$_\odot/yr$ \citep{BouGrAl2007}.% (Bouvier et al. 2007).

%Кроме того, у CTTS наблюдается интенсивный ветер. Темп потери массы, в среднем,  на порядок величины меньше, чем темп аккреции.  Рассматривались различные модели ветра: дисковый ветер, ускоряемый магнитной центрифугой вращающегося диска \citep{Shu1994, Ustyugova2006, Rom2015},% (Shu et al. 1994; Ustyugova et al. 2006; Romanova  Owocki 2016),  
%~конический ветер, стартующий на границе аккреционного  диска  и магнитосферы, и ускоряемый магнитным давлением \citep{Rom2009}, %(Romanova et al. 2009), 
%~полярный ветер, ускоряемый МГД волнами, которые возникают при аккреции газа на поверхность звезды \citep{Cran2008}, %(Cranmer 2008),  
%~и другие.

In addition, CTTSs exhibit a strong wind. The mass loss rate is, on average, an order of magnitude smaller than the accretion rate. Various wind models have been considered: a disk wind accelerated by the magnetic centrifuge of a rotating disk \citep{Shu1994, Ustyugova2006, Rom2015},%(Shu et al. 1994; Ustyugova et al. 2006; Romanova & Owocki 2016), 
~a conical wind starting at the accretion disk-magnetosphere boundary and accelerated by magnetic pressure \citep{Rom2009},%(Romanova et al. 2009),
~a polar wind accelerated by MHD waves that arise during gas accretion onto the stellar surface \citep{Cran2008},%(Cranmer 2008),
~and others.

%Анализ эмиссионных спектров CTTS показывает, что как аккреционный канал, так и область ветра неоднородны: в аккреционном канале есть струи повышенной плотности, а в ветре  --  облака большей плотности  по сравнению с окружающими потоками \citep{Fish2008, KwFi2011}. %(Fischer et al. 2008,  Kwan Fischer 2011). 
%~На больших расстояниях ветер коллимирован магнитным полем в биполярные джеты  (см. обзор \cite{PCE2023}). % Pascucci et al. 2023).

Analysis of CTTS emission spectra shows that both the accretion flows and the wind region are heterogeneous: the accretion flows contain  streams  of increased density, while the wind contains clouds of higher density compared to the surrounding flows \citep{Fish2008, KwFi2011}. %(Fischer et al. 2008; Kwan & Fischer 2011).
~At large distances, the wind is collimated by the magnetic field into bipolar jets (see review by \cite{PCE2023}). %Pascucci et al. 2023).

%Обзоры наблюдаемых характеристик  CTTS можно найти в публикациях: \cite{HarHCal2016}, %Hartmann et al. (2016), 
%~\cite{BouGrAl2007}, %Bouvier et al. (2007), 
%~\cite{Petrov2021}. %Petrov (2021).  

Reviews of the observed characteristics of CTTS can be found in \cite{HarHCal2016}, %Hartmann et al. (2016), 
~\cite{BouGrAl2007}, %Bouvier et al. (2007), 
~\cite{Petrov2021}. %Petrov (2021).  Hartmann et al. (2016), Bouvier et al. (2007), and Petrov (2021).

%Спектральные наблюдения  CTTS позволяют  “увидеть” потоки газа и измерить скорости падения и  истечения по профилям спектральных линий, уширенных эффектом Допплера, и проследить как меняется динамика этих процессов на разных временных шкалах -- от суток до лет. С этой целью проводятся мониторинги избранных объектов. Наиболее плотные ряды наблюдений AA Tau были проведены одновременно на нескольких обсерваториях \citep{Bou2003}. %(Bouvier et al. 2003). 
%~Серии спектральных и фотометрических наблюдений CTTS, охватывающие несколько лет, проводились для звезд:  SU~Aur -- \cite{Gia1993, BasJ1995, Petrov1996, PetrovGr2019}; %(Giampapa et al. 1993; JohnsBasri 1995; Petrov et al. 1996; 2019),  
%~RY~Tau -- \cite{BaArP2016, PetrovGr2019}; % (Babina et al. 2016; Petrov et al, 2019), 
%~RW~Aur -- \cite{AlBasC2005, Petrov1996, Tak2016}; %(уточнить ссылки: Petrov et al. 1996; Alencar et al. 2005; Takami et al. 2016) ,  
%~DR~Tau -- \cite{AlJKBas2001}. %(Alencar et al. 2001). 

Spectral observations of CTTSs allow us to “visualize” gas flows and measure infall and outflow velocities using Doppler-broadened spectral line profiles, and track how the dynamics of these processes change over different time scales-from days to years. For this purpose, monitoring of selected objects is being conducted. The most dense series of observations of CTTS AA~Tau were conducted simultaneously at several observatories \citep{Bou2003}. %(Bouvier et al. 2003). 
~Spectroscopic and photometric observations of CTTSs spanning several years were conducted for the stars SU~Aur -- \cite{Gia1993, BasJ1995, Petrov1996, PetrovGr2019}; %(Giampapa et al. 1993; JohnsBasri 1995; Petrov et al. 1996; 2019),  
~RY~Tau -- \cite{BaArP2016, PetrovGr2019}; % (Babina et al. 2016; Petrov et al, 2019), 
~RW~Aur -- \cite{AlBasC2005, Petrov1996, Tak2016}; %(уточнить ссылки: Petrov et al. 1996; Alencar et al. 2005; Takami et al. 2016) ,  
~DR~Tau -- \cite{AlJKBas2001}. %(Alencar et al. 2001).%(Giampapa et al. 1993; Johns & Basri 1995; Petrov et al. 1996; 2019), RY Tau (Babina et al. 2026; Petrov et al. 2019), RW Aur (Petrov et al. 1996; Alencar et al. 2005; Takami et al. 2016), and DR Tau (Alencar et al. 2001).

%В данной статье мы представляем результаты 11-летней серии наших наблюдений  RY~Tau,  с 2013 по 2024~гг. Цель этого исследования  -- определение временных характеристик и выявление причинно-следственных связей в событиях аккреции и ветра. RY Tau -- одна из самых ярких и часто наблюдаемых CTTS северного неба. Основные параметры звезды, согласно \cite{CalMuz2004}: % Calvet et al. (2004),  
%~T$_e$=5945$\pm$142~K, L$_*$=9.6$\pm$1.5~L$_\odot$,  M$_*$=2.0$\pm$0.3~M$_\odot$, R$_*$=2.9$\pm$0.4~R$_\odot$.

In this paper, we present the results of our 11-year series of observations of RY~Tau, from 2013 to 2024. The goal of this study is to determine the timing characteristics and identify causal relationships in accretion and wind events. RY~Tau is one of the brightest and most frequently observed CTTSs in the northern sky. The main parameters of the star, according to \cite{CalMuz2004}, %Calvet et al. (2004), 
~are T$_e$=5945$\pm$142~K, L$_*$=9.6$\pm$1.5~L$_\odot$,  M$_*$=2.0$\pm$0.3~M$_\odot$, R$_*$=2.9$\pm$0.4~R$_\odot$.

%Звезда видна под большим углом наклона оси вращения к лучу зрения, тем не менее не удается заметить какой-либо периодической переменности, вызванной осевым вращением звезды \citep{Zajtseva2010}.%(Zajtseva 2010).  
%~Вероятно, магнитный диполь не наклонен к оси вращения звезды.  Проекция скорости вращения $v\,\sin i$, определяемая по фотосферным линиям, составляет  52$\pm$2~км\,с$^{-1}$ \citep{Bou1990, PetrovZaj1999}.% (Bouvier 1990; Petrov et al. 1999).  
%~Зная радиус звезды,  проекцию скорости вращения  $v\,\sin i$ и угол наклона $i$, можно оценить период вращения: P$_*$=2.84$\pm$0.40 $^{d}$ , то есть RY~Tau относится к быстро вращающимся CTTS. Соответствующий радиус коротации в протопланетном диске RY~Tau составляет 10$\pm$1R$_\odot$.

The star is visible at a large inclination of its rotation axis to the line of sight; however, no periodic variability caused by the star's axial rotation can be observed \citep{Zajtseva2010}.%(Zajtseva 2010). 
~The magnetic dipole is likely not tilted to the star's rotation axis. The projected rotation velocity $v\,\sin i$, determined from photospheric lines, is 52$\pm$2~km\,s$^{-1}$ \citep{Bou1990, PetrovZaj1999}.% 52 ± 2 km/s (Bouvier 1990; Petrov et al. 1999). 
~Knowing the star's radius, the projected rotation velocity $v\,\sin i$, and the tilt angle $i$, we can estimate the rotation period: P$_*$=2.84$\pm$0.40~$^{d}$, i.e., RY~Tau belongs to the rapidly rotating CTTS. The corresponding corotation radius in the protoplanetary disk of RY~Tau is 10$\pm$1R$_\odot$. 

%Протопланетный диск RY~Tau хорошо разрешен интерферометром ALMA, ось вращения  диска  наклонена под углом 65\,$^{\circ}$ к лучу зрения \citep{LHH2019}.% (Long et al. 2019). 
%~Наклон внутреннего диска, определенный по интерферометрии в К-диапазоне, $i$=60\,$^{\circ}$ \citep{Perraut2021}.% (Perraut et al. 2021).  
%~При угле наклона 60-65\,$^{\circ}$ луч зрения пересекает дисковый ветер и магнитосферу звезды. Наблюдаемая фотометрическая переменность RY~Tau, вероятно, в значительной мере обусловлена поглощением в запыленном дисковом ветре \citep{BaArP2016, Dav2020}.% (Babina et al. 2016; Davies et al. 2020). 

The protoplanetary disk of RY~Tau is well resolved by the ALMA interferometer; the disk rotation axis is tilted at 65\,$^{\circ}$ to the line of sight \citep{LHH2019}.%(Long et al. 2019). 
~The tilt of the inner disk, determined from K-band interferometry, $i$=60\,$^{\circ}$ \citep{Perraut2021}.%(Perraut et al. 2021). 
~At a tilt angle of 60-65\,$^{\circ}$, the line of sight intersects the disk wind and the star's magnetosphere. The observed photometric variability of RY~Tau is likely largely due to absorption in the dusty disk wind \citep{BaArP2016, Dav2020}.%(Babina et al. 2016; Davies et al. 2020).

%\textbf{Нерегулярные увеличения блеска происходили в 1983-1984 и 1996-1997 годах  \citep{Herb1984, Herb1994, Zaj1996}.%(Herbst & Stine 1984; Herbst et al. 1994; Zajtseva et al. 1996). 
%~Наиболее продолжительный ряд фотометрических наблюдений RY~Tau с 1965 по 2000 год был проанализирован Зайцевой (2010): обнаружены квазипериодические изменения блеска,  вызванные, предположительно,  затмениями пылевым облаком в околозвездном диске. Периодичности, связанной с осевым вращением звезды, обнаружено не было.}

Irregular brightness increases occurred in 1983–1984 and 1996–1997 \citep{Herb1984, Herb1994, Zaj1996}.%(Herbst & Stine 1984; Herbst et al. 1994; Zajtseva et al. 1996).
~The longest series of photometric observations of RY Tau, from 1965 to 2000, was analyzed by Zajtseva (2010): quasi-periodic brightness variations were detected, presumably caused by eclipses by a dust cloud in the circumstellar disk. No periodicity associated with the axial rotation of the star was detected.

%\textbf{По фотометрическим наблюдениям RY~Tau в 1983–2004 гг. обнаружен период 377$\pm$10 суток. Высказано предположение, что период вызван присутствием формирующихся тел в протопланетном диске \citep{Ism2012}. %(N. Z. Ismailov and A. N. Adygezalzade (2012, Astronomy Reports, 2012, Vol. 56, No. 2, pp. 131–137).
%~Период 23 суток был обнаружен в изменениях эмиссионной линии MgII 2800~\AA\ в ультрафиолетовом спектре RY~Tau \citep{Ism2015}.% (Ismailov et al. ,2015). 
%~Близкие значения периода были обнаружены также и в сериях  фотометрических наблюдений \citep{Bou1993, Gahm93a}.} %(Bouvier et al. 1993; Gahm et al. 1993a).

Photometric observations of RY Tau from 1983–2004 revealed a period of 377$\pm$10 days. It has been suggested that this period is caused by the presence of forming bodies in the protoplanetary disk \citep{Ism2012}. %. (N. Z. Ismailov and A. N. Adygezalzade (2012, Astronomy Reports, 2012, Vol. 56, No. 2, pp. 131–137).
~A period of 23 days was detected in variations of the MgII 2800~\AA\ emission line in the ultraviolet spectrum of RY~Tau \citep{Ism2015}.%(Ismailov et al. 2015).
~Similar period values were also found in a series of photometric observations \citep{Bou1993, Gahm93a}.%(Bouvier et al. 1993; Gahm et al. 1993a).

%~Кроме того,  по данным наблюдений 2013-2020~гг. были замечены  изменения потока в линии Н$\alpha$ на лучевой скорости  RV=$-95\pm$5 км\,с$^{-1}$, с периодом 21.6~дня,  что было интерпретировано как струи плотности в дисковом ветре, вызванные неоднородной структурой протопланетного диска на расстоянии $\sim$0.2~а.е. от звезды \citep{PetrovRo2021, Petrov2023}. %(Petrov et al. 2021, 2023).

In addition, observations from 2013 to 2020 revealed variations in the Н$\alpha$ flux at a radial velocity of RV=$-95\pm$5 km\,s$^{-1}$ with a period of 21.6~days. These were interpreted as density streams in the disk wind caused by the inhomogeneous structure of the protoplanetary disk at a distance of $\sim$0.2~AU from the star \citep{PetrovRo2021, Petrov2023}. %(Petrov et al. 2021, 2023).

%У RY~Tau наблюдается протяженный джет, в котором видны узлы с  динамическим возрастом менее 10 лет \citep{StOnge2008, AADC2009, Skinner2018, Tak2023}. %(St-Onge  Bastien, 2008;  Agra-Amboage et al. 2009; Skinner et al. 2018; Takami et al. 2023;).  
%~Джет искривлен, что может быть следствием наклона аккреционного диска, вызванного присутствием планеты или мало-массивного компонента звезды \citep{GarPo2019}. %(Garufi et al., 2019). 
%~RY~Tau является источником  переменного рентгеновского излучения, указывающим на присутствие горячей плазмы с температурой T$\approx$50~MK \citep{Skinner2016}. %(Skinner et al. 2016).

RY~Tau exhibits an extended jet containing knots with a dynamical age of less than 10 years \citep{StOnge2008, AADC2009, Skinner2018, Tak2023}.%(St-Onge & Bastien, 2008; Agra-Amboage et al. 2009; Skinner et al. 2018; Takami et al. 2023). 
~The jet is warped, which may be due to the tilt of the accretion disk caused by the presence of a planet or a low-mass component of the star \citep{GarPo2019}. %(Garufi et al., 2019). 
~RY~Tau is a source of variable X-ray emission, indicating the presence of hot plasma with a temperature of T$\approx$50~MK \citep{Skinner2016}. % (Skinner et al., 2016).

\section{Observations}
%Основной массив наших спектральных наблюдений RY Tau был получен в Крымской астрофизической обсерватории (КрАО РАН) на 2.6-м телескопе ЗТШ  с помощью  спектрографа ЭСПЛ \citep{LagPlach2020}.% (Лагутин и др. 2020).
%~Спектральное разрешение составляло $\lambda/\Delta\lambda\approx$27000  при ширине входной щели 2 угловых секунды. Анализировались участки спектра в области линий Н$\alpha$ и D~NaI. 

%Спектральные наблюдения проводились также в Коуровской обсерватории УрФУ на 1.21-м телескопе ($\lambda/\Delta\lambda\approx$27000)  и, в один из сезонов (2015-2016~гг.), на 2.5-м телескопе NOT, $\lambda/\Delta\lambda\approx$15000, и на 2.2-м телескопе CAHA, $\lambda/\Delta\lambda\approx$30000 (Испания).  
	
%Несколько спектров были получены на 2.4-м телескопе TNT (Тайланд) и на 2.2-м телескопе  САНА  (Испания).
%\textbf{Поскольку мы измеряем довольно широкие детали профиля спектральных линий (см. следующий раздел), различие в спектральном разрешении спектров, полученных на разных инструментах, было не существенным.}

The bulk of our spectral observations of RY~Tau were carried out at the Crimean Astrophysical Observatory (CrAO RAS) using the 2.6-m ZTSh telescope and the ESPL spectrograph \citep{LagPlach2020}.%(Lagutin et al. 2020). 
~The spectral resolution was $\lambda/\Delta\lambda\approx$27000 with an entrance slit width of 2 arcseconds. Spectral regions in the Н$\alpha$ and D~NaI lines were analyzed. 
Spectral observations were also conducted at the Kourovka Observatory of UrFU using the 1.21-m telescope ($\lambda/\Delta\lambda\approx$27000) and, during one season (2015-2016), using the 2.5-m NOT telescope, $\lambda/\Delta\lambda\approx$15000, and the 2.2-m CAHA telescope, $\lambda/\Delta\lambda\approx$30000 (Spain).

Since we measure fairly broad spectral line profile details (see the next section), the difference in spectral resolution between the different instruments was insignificant.

%Фотометрические наблюдения RY~Tau в системе BVRI проводились в те же ночи, что и спектральные, на телескопах Крымской астрофизической обсерватории, в основном на  1.25-м телескопе АЗТ-11. Несколько оценок блеска было получено на 0.8-м телескопе РК-800. Типичная точность фотометрических измерений блеска звезды составляла $\pm$0.02$^{m}$ в каждой полосе. В тех случаях, когда погодные условия не позволяли выполнить точную фотометрию, мы использовали базу данных AAVSO  (\emph{https://www.aavso.org}). Следует отметить хорошее согласие (в пределах $\pm$0.02$^{m}$) между нашими оценками блеска и фотометрическими данными из базы данных AAVSO. 

Photometric observations of RY~Tau in the BVRI system were conducted on the same nights as the spectroscopic observations using telescopes of the Crimean Astrophysical Observatory, primarily the 1.25-m AZT-11 telescope. Several brightness estimates were obtained using the 0.8-m RK-800 telescope. The typical accuracy of photometric measurements of the star's brightness was $\pm$0.02$^{m}$ in each band. In cases when weather conditions prevented precise photometry, we used the AAVSO database (\emph{https://www.aavso.org}). In general, good agreement (within $\pm$0.02$^{m}$) between our brightness estimates and the photometric data from the AAVSO database should be noted.

%Более подробное описание  наблюдений и анализ полученных результатов были изложены в наших предыдущих публикациях: \cite{BaArP2016, PetrovGr2019, PetrovRo2021, Petrov2023}.% (Babina et al. 2016; Petrov et al. 2019; Petrov et al. 2021; Petrov et al. 2023).  
%~В общей сложности, наши наблюдения RY~Tau охватывают 220 ночей за 11 сезонов с 2013 по 2024~гг.

A more detailed description of the observations and analysis of the results were presented in our previous publications (\cite{BaArP2016, PetrovGr2019, PetrovRo2021, Petrov2023}).%Babina et al. 2016; Petrov et al. 2018; Petrov et al. 2021; Petrov et al. 2023). 
~In total, our observations of RY~Tau cover 220 nights over 11 seasons from 2013 to 2024.

\section{Observation Analysis}
%При анализе динамики аккреции и ветра на луче зрения мы использовали эмиссионную линию Н$\alpha$ и абсорбционные линии дублета D~NaI. Ограниченный размер детектора изображения в спектрографе ЭСПЛ не позволял регистрировать одновременно и другие характерные индикаторы аккреции и ветра,  такие как HeI 5876~\AA\ или линии инфракрасного триплета CaII. На Рис.1  показаны наиболее характерные профили линий Н$\alpha$ и D~NaI, расположенные (сверху вниз) в порядке убывания признаков ветра и усиления признаков аккреции. На всех спектрах, показанных в данной статье, шкала длин волн и лучевых скоростей -- астроцентрическая. Лучевая скорость RY~Tau принята равной +18~км\,с$^{-1}$. Даты наблюдений указаны в формате  HJD-2450000. Шкала лучевой скорости в области линий дублета NaI приведена относительно линии D1.

We used the Н$\alpha$ emission line and the D absorption lines of the NaI doublet to analyze the accretion and wind dynamics along the line of sight. The limited size of the imaging detector in the ESPL spectrograph prevented us from simultaneously recording other accretion and wind indicators, such as HeI 5876~\AA\ or the infrared CaII triplet lines. Figure 1 shows the most characteristic Н$\alpha$ and D line profiles, arranged (from top to bottom) in order of decreasing wind signatures and increasing accretion signatures. All spectra shown in this article are displayed on an astrocentric wavelength and radial velocity scale. The radial velocity of RY~Tau is assumed to be +18~km\,s$^{-1}$. Observation dates are given in the HJD-2450000 format. The radial velocity scale in the NaI doublet line region is given relative to the D1 line.

%Эмиссия Н$\alpha$ образуется как в магнитосфере, так и в потоках ветра \citep{Muz2001, KwFi2011}.% (Muzerolle et al. 2001, Kwan Fischer 2011). 
%~Профили бальмеровских линий водорода в разных моделях ветра CTTS исследовались в работе \cite{KuRom2011}. 
%Kurosawa et al (2011). 
%~Характерная депрессия профиля Н$\alpha$ в коротковолновом крыле линии  является индикатором поглощения в ветре на луче зрения. В спектрах RY~Tau эта депрессия  временами смещается к центру линии и даже в длинноволновое крыло, что указывает на поглощение в падающем на звезду потоке газа в моменты усиления  аккреции  (см. Рис.1, HJD~8410 и 8414).

Н$\alpha$ emission is formed both in the stellar magnetosphere and in wind streams \citep{Muz2001, KwFi2011}.%(Muzerolle et al. 2001, Kwan & Fischer 2011). 
~The profiles of the hydrogen Balmer lines in different CTTS wind models were studied in \cite{KuRom2011}. %Kurosawa et al. (2011). 
~The characteristic depression of the Н$\alpha$ profile in the shortwave wing of the line is an indicator of absorption in the line-of-sight wind. In the spectra of RY~Tau, this depression sometimes shifts toward the line center and even into the longwave wing, indicating absorption in the gas flow infalling onto the star during periods of enhanced accretion (see Fig. 1, HJD 8410 and 8414).

%Абсорбционные линии резонансного дублета D~NaI образуются как в потоках падающего газа внутри магнитосферы, так и  в наиболее плотных областях ветра \citep{CalMuz2004}.% (Muzerolle et al. 2001). 
%~Абсорбция в “синем” крыле линий D~NaI, характеризующем поглощение в ветре,  простирается  до  скоростей более 200~км\,с$^{-1}$.  В линии D1 “красное” крыло частично блендировано теллурическими линиями воды, а в более насыщенной линии D2 в те даты наблюдений,  когда признаки ветра были минимальны, “красное” крыло  простиралось до скоростей $\sim250$~км\,с$^{-1}$ (см. Рис.1), что характеризует скорость падения газа.  По максимальной скорости падения можно оценить радиус магнитосферы Rm (см., например, Takasao et al. 2022).% Takasao et al. 2022). 
%~С учетом неопределенностей массы и радиуса RY~Tau,  R$_{m}$=10$\pm$4~R$_\odot$.  

Absorption lines of the D~NaI resonance doublet are formed both in infalling gas streams within the magnetosphere and in the densest regions of the wind \citep{CalMuz2004}.%(Muzerolle et al. 2001). 
~Absorption in the “blue” wing of the D~NaI lines, which characterizes absorption in the wind, extends to velocities exceeding 200~km\,s$^{-1}$. In the D1 line, the “red” wing is partially blended with telluric water lines, and in the more saturated D2 line, on observation dates when wind signatures were minimal, the “red” wing extended to velocities of $\sim250$~km\,s$^{-1}$ (see Fig. 1), which characterizes the gas infall velocity. The maximum infall velocity can be used to estimate the magnetospheric radius Rm (see, e.g., Takasao et al. 2022). Taking into account the uncertainties in the mass and radius of RY~Tau, R$_{m}$=10$\pm$4~R$_\odot$.

\begin{figure}
\includegraphics[scale=0.54]{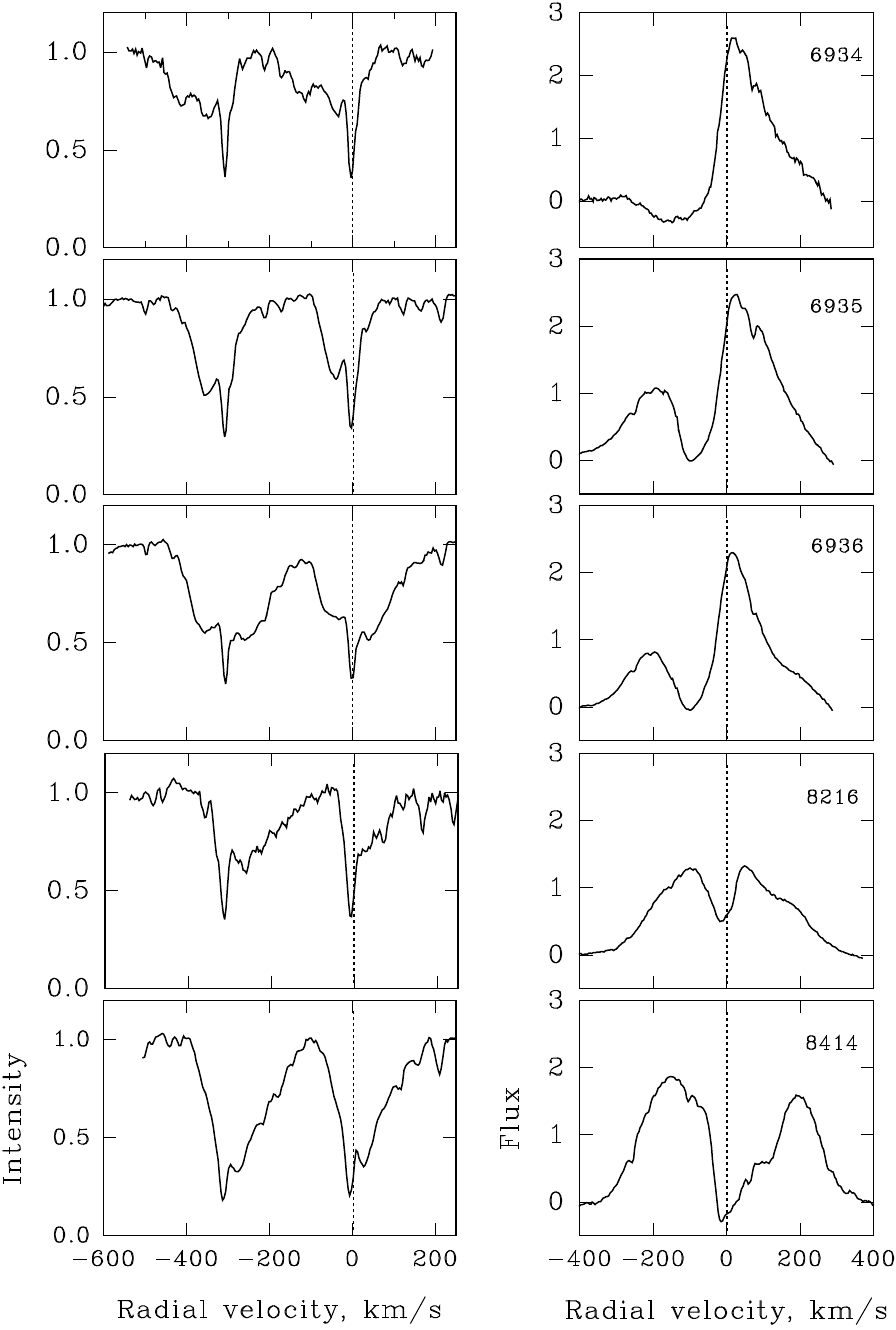}
\caption{Examples of NaI~D (left panel) and H$\alpha$ (right panel) line profiles. Observation dates (HJD-2450000) are indicated in the upper right corner of the right panel. The H$\alpha$ line flux is given in units of 3.67·10$^{-13}$ erg\,cm$^{-2}$s$^{-1}$.}
\label{fig:a1}
\end{figure}

%Следует учесть, что в условиях магнитосферы атомы натрия почти полностью ионизованы, поэтому профиль абсорбции в линиях  D~NaI отражает динамику наиболее плотных областей, где доля нейтральных атомов еще достаточно велика. Скорость,  измеренную по протяженности “красного” крыла этих линий, можно считать нижней границей скорости падения газа на поверхность звезды.

It should be noted that under magnetospheric conditions, sodium atoms are almost completely ionized, so the absorption profile in the D~NaI lines reflects the dynamics of the densest regions, where the fraction of neutral atoms is still quite high. The velocity measured from the extent of the “red” wing of these lines can be considered the lower limit of the velocity of gas infall onto the star's surface.

%В профилях линий D~NaI присутствует также узкая абсорбция межзвездного поглощения на лучевой скорости около -10~км\,с$^{-1}$ относительно звезды.

The D~NaI line profiles also contain a narrow interstellar absorption band at a radial velocity of about -10~km\,s$^{-1}$ relative to the star.

%При анализе переменности профилей линий Н$\alpha$ и D~NaI мы выбрали следующие параметры (см. Рис. 2):  Fb и Fr -- \textit{потоки} излучения в “синей” и  “красной” половине профиля, относительно нулевой лучевой скорости. Поток излучения вычислялся по эквивалентной ширине линии EW и звездной величине V в момент наблюдения:
%F=EW(\AA)\cdot 10$^{-0.4(V-10)}$,
%в единицах  3.67·10$^{-13}$ эрг\,cм$^{-2}$с$^{-1}$}.

When analyzing the variability of the Н$\alpha$ and D~NaI line profiles, we chose the following parameters: (see Fig. 2) Fb and Fr are the radiative fluxes in the “blue” and “red” halves of the profile, relative to zero radial velocity. The radiative flux was calculated from the equivalent line width EW and the stellar magnitude V at the time of observation:
F=EW(\AA)$\cdot~10^{-0.4(V-10)}$,
in units of   3.67·10$^{-13}$ erg\,cm$^{-2}$s$^{-1}$.

%За единицу измерения принят поток в континууме звезды 10-й величины в области полосы V. Более уместно было бы использовать фотометрическую величину R,  но для некоторых дат наблюдений в базе данных AAVSO была доступна только величина V. Использование величины V при определении потока в линии Н$\alpha$ дает систематическую погрешность не более 10\%, так как при изменении блеска RY~Tau в пределах V=9.6-11.0$^{m}$ цвет (V-R) остается почти постоянным: 1.1\pm0.1$^{m}$ \citep{PetrovGr2019}.

The unit of measurement is the continuum flux of a 10th-magnitude star in the V band. It would be more appropriate to use the photometric magnitude R, but for some observation dates, only the magnitude V was available in the AAVSO database. Using the magnitude V to determine the Н$\alpha$ flux yields a systematic error of no more than 10\%, since as the brightness of RY~Tau varies within the range of V=9.6-11.0$^{m}$, the color (V-R) remains almost constant: $1.1\pm0.1^{m}$ \citep{PetrovGr2019}.

%Изменения потока Fb отражают изменения плотности ветра на луче зрения.  Параметр Fr характеризует  излучение магнитосферы и ветра в линии Н$\alpha$.

%В профиле линии D1~NaI мы измеряли эквивалентную ширину “синего” и “красного” абсорбционных крыльев линии D1, протяженностью 1.5~\AA (76~км\,с$^{-1}$) каждый, исключая центральный участок, где находится узкая межзвездная абсорбция (см. Рис. 2)

Changes in the Fb flux reflect changes in the wind density along the line of sight. The parameter Fr characterizes the emission of the magnetosphere and wind in the Н$\alpha$ line. In the D1~NaI line profile, we measured the equivalent width of the “blue” and “red” absorption wings of the D1~line, each 1.5~\AA (76~km\,s$^{-1}$) long, excluding the central region where the narrow interstellar absorption is located (see Fig. 2).  

\begin{figure}
\includegraphics[scale=0.55]{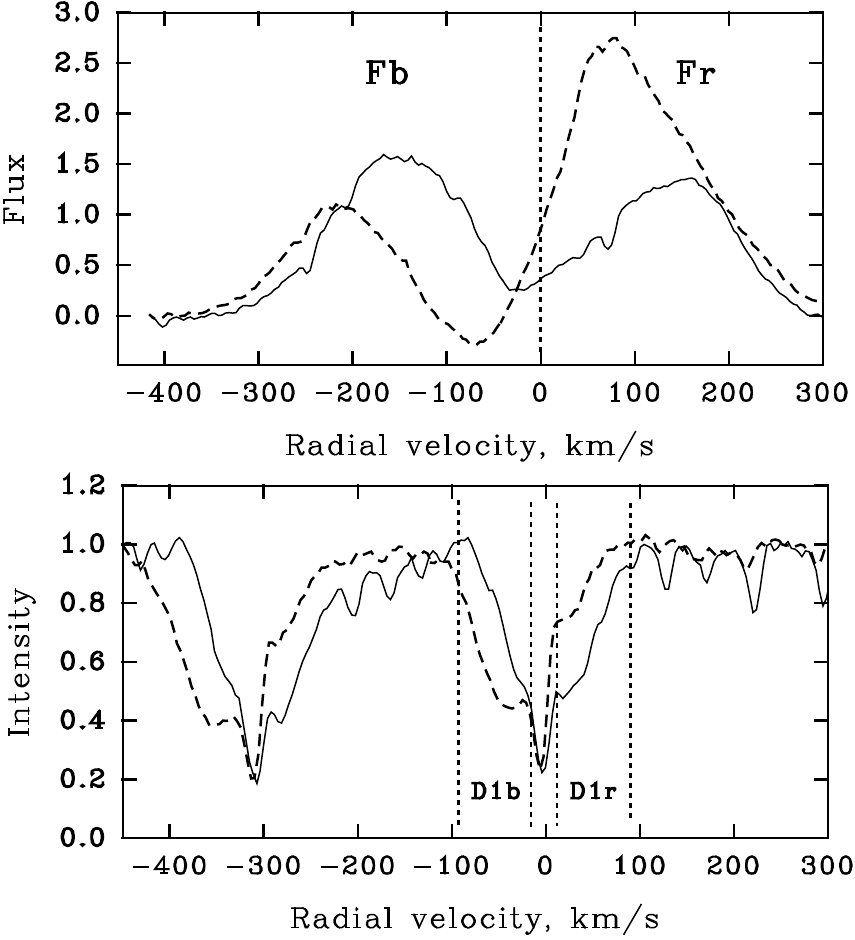}
\caption{Profiles of the Н$\alpha$ and D~NaI lines. Observation times: HJD~8734 (solid line) and 8765 (dashed line).}
\label{fig:a2}
\end{figure}

%Параметр D1r характеризует поглощение в падающем на звезду потоке внутри магнитосферы, а параметр D1b -- поглощение в ветре за пределами магнитосферы, вероятно в основании ветра или в наиболее плотных струях или облаках ветра.          

%На Рис.3 показано, как изменялись со временем блеск звезды V, полный поток излучения в линии Н$\alpha$, и абсорбция D1r, характеризующая аккрецию.  Во время минимальной аккреции (HJD~7650--7657) наблюдался минимальный поток в Н$\alpha$ и минимальный блеск V.  Наблюдения проводились 7 ночей подряд, и все время в профилях спектральных линий были признаки сильного ветра со скоростью до 200~км\,с$^{-1}$  и не было признаков аккреции (EW~D1r~$\approx$0).  

The D1r parameter characterizes the absorption in the magnetospheric flux incident on the star, while the D1b parameter characterizes the absorption in the wind outside the magnetosphere, likely at the wind base or in the densest wind streams or clouds.

Figure 3  shows how the star's V brightness, total Н$\alpha$ flux, and D1r absorption, which characterizes accretion, changed over time. During minimal accretion (HJD 7650 -- 7657), the minimum Н$\alpha$ flux and minimum V brightness were observed. Observations were performed over seven consecutive nights, and throughout the entire period, the spectral line profiles showed signs of a strong wind with velocity up to 200~km\,s$^{-1}$ and no signs of accretion (EW~D1r~$\approx$0).

%В другие интервалы непрерывных наблюдений происходили заметные изменения потоков аккреции и ветра на шкале времени  от суток и более (см., например, Рис. 1).  

%Следует отметить, что в изменениях указанных параметров спектральных линий (Fb, Fr, D1b, D1r) \textit{не прослеживается} период вращения звезды. В некоторых интервалах непрерывных наблюдений в течение 6 суток (около двух оборотов звезды) не было существенных изменений профилей линий.  Это означает, что потоки газа -- аккрецирующего или истекающего -- \textit{аксиально симметричны}.

%В даты наблюдений HJD~8215 и 8216 (Рис. 1) а в профиле линий D~NaI видны наиболее сильные признаки аккреции. В профиле Н$\alpha$ в эти даты была видна абсорбция в центре линии. В другой момент времени, HJD~8414, абсорбция явно сместилась в “красное” крыло линии Н$\alpha$, указывая на падение газа на звезду. 

During other continuous observation intervals, noticeable changes in the accretion and wind fluxes occurred on time scales of a day or longer (see, for example, Fig. 1).

It should be noted that the variations in the aforementioned spectral line parameters (Fb, Fr, D1b, D1r) do not reveal the star's rotation period. During some continuous observation intervals lasting 6 days (approximately two stellar rotations), there were no significant changes in the line profiles. This indicates that the gas fluxes -- accreting or outflowing -- are \textit{axially symmetric}.
On the observation dates of HJD~8215 and 8216 (Fig. 1), the strongest signs of accretion are visible in the D~NaI line profile. The Н$\alpha$ profile on these dates showed absorption at the line center. At another day, HJD~8410, the absorption is clearly shifted to the “red” wing of the Н$\alpha$ line, indicating gas infall onto the star.

\begin{figure}
\includegraphics[scale=0.51]{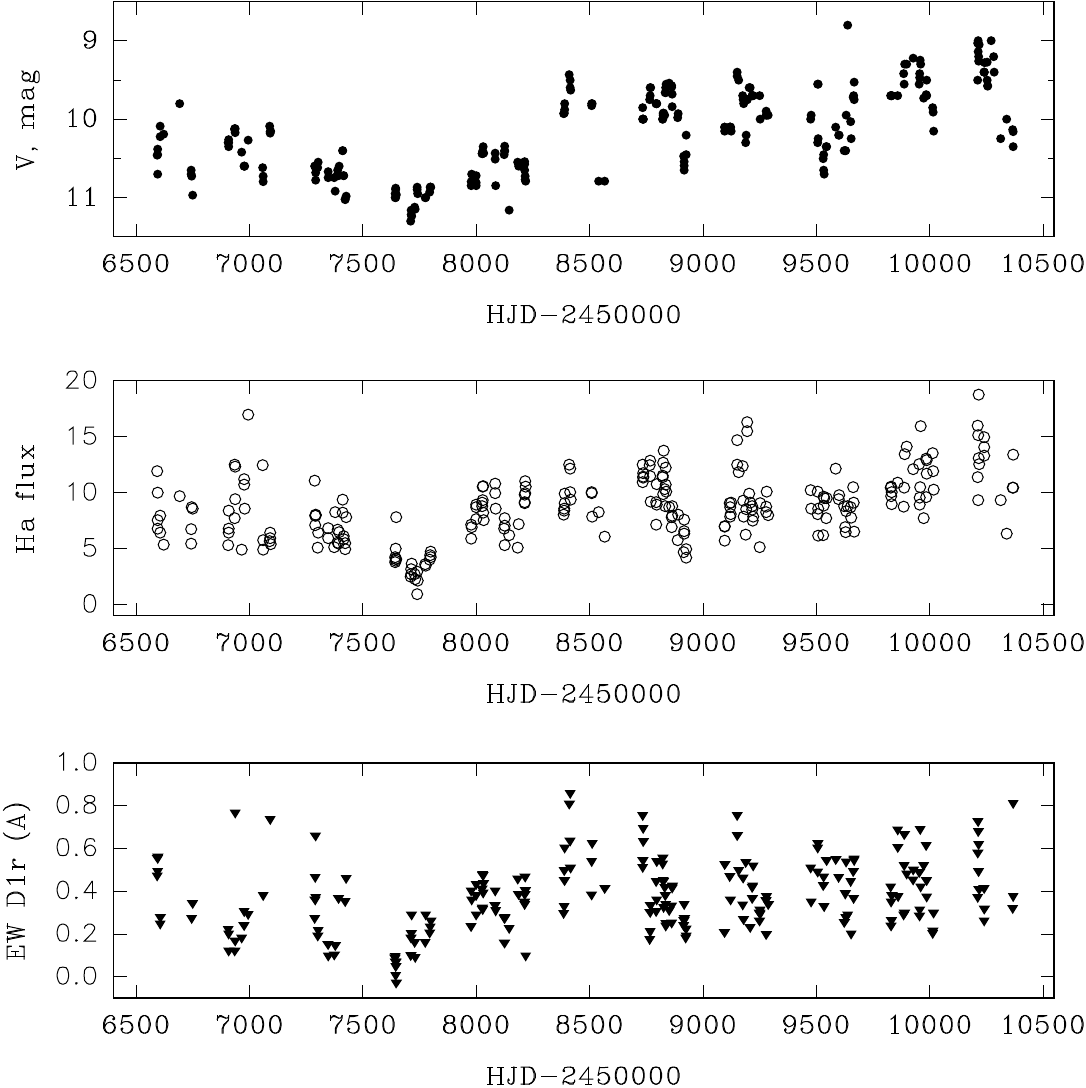}
\caption{Variations in the star's V-band brightness, Н$\alpha$ flux, and accretion flow absorption (EW~D1r) over 11 years.}
\label{fig:a3}
\end{figure}
%Анализ спектров показал, что среди четырех параметров -- Fb, Fr, D1b и D1r -- корреляция наблюдается только между двумя параметрами, D1r (аккреция) и Fb (ветер): при усилении аккреции (D1r) скорость ветра уменьшается,  депрессия в синем крыле Н$\alpha$ смещается к центру линии (см. Рис.1), поэтому увеличивается поток Fb -- аккреция “гасит ветер (Рис. 4).

Of the four parameters -- Fb, Fr, D1b and D1r -- a correlation is observed only between two parameters, D1r (accretion) and Fb (wind): as accretion (D1r) increases, the wind speed decreases, and the depression in the blue Н$\alpha$ wing shifts toward the line center (see Fig. 1), resulting in an increase in the Fb flux -- accretion dampens the wind (Fig. 4).

\begin{figure}
\includegraphics[scale=0.77]{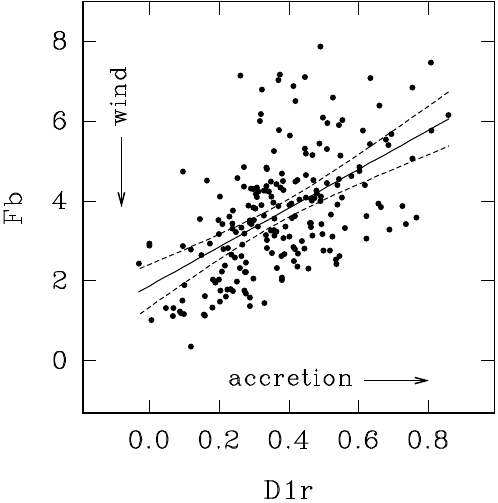}
\caption{Correlation between the absorption in the accreting flow (D1r) and the emission flux (Fb) in the "blue" wing of the Н$\alpha$ emission line. The dashed lines indicate the 99\% confidence interval. Low Fb flux values correspond to high wind density along the line of sight. As accretion increases, the wind density decreases and the Fb flux increases.}
\label{fig:a4}
\end{figure}

\section{Time series cross-correlation}
%Мы анализируем переменность наблюдаемых параметров: признак аккреции (D1r) и признак ветра (Fb). Эти признаки присутствуют в каждом спектре, они переменны на шкале времени в несколько дней и между ними есть заметная корреляция. Но мы не знаем  $\textit{a~priori}$, одновременно ли меняются признаки аккреции и ветра, или изменения в одной спектральной линии происходят на день-два раньше, чем в другой. Например, сначала появляется признак аккреции в линиях D~NaI, а потом изменяются признаки ветра в линии Н$\alpha$.  Проверим эту гипотезу: сначала аккреция, потом  ветер. 

We analyze the variability of observed parameters: the accretion signature (D1r) and the wind signature (Fb). These signatures are present in each spectrum, vary on a time scale of several days, and are significantly correlated. However, we do not know \textit{a~priori} whether the accretion and wind signatures change simultaneously, or whether changes in one spectral line occur a day or two earlier than in the other. For example, the accretion signature in the D~NaI lines appears first, followed by changes in the wind signature in the Н$\alpha$ line. We will test this hypothesis: first accretion, then wind.

%Обозначим переменные величины: D1r -- смещенная в красную сторону абсорбция в резонансной линии D1~NaI. Эквивалентная ширина этой абсорбции измеряется в Ангстремах. Другая величина -- поток излучения Fb в коротковолновом крыле эмиссионного профиля Н$\alpha$. %выраженный в единицах потока в континууме звезды V=10$^m$ [3.67·10$^{-13}$ эрг\,cм$^{-2}$с$^{-1}$]. 

We denote the variables: D1r is the redshifted absorption in the D1~NaI resonance line. The equivalent width of this absorption is measured in Angstroms. Another quantity is the radiation flux Fb in the short-wavelength wing of the Н$\alpha$ emission profile.% expressed in units of flux in the continuum of the star V=10m [3.67‧10-13 erg/cm2 ‧ s].

%Далее, мы используем метод кросс-корреляции временных рядов. Создаем \textit{равномерный} ряд данных [HJD, D1r] с шагом один день, и такой же ряд данных [HJD, Fb]. Даты, в которые не было наблюдений, содержат пустые ячейки ряда. Строим график зависимости Fb от D1r и определяем коэффициент корреляции. Затем сдвигаем на один день ряд [HJD, D1r] относительно ряда [HJD,Fb], так чтобы параметр Fb сравнивался с параметром D1r в предыдущий день, снова строим график и определяем коэффициент корреляции. Таким образом мы проверяем, изменится ли корреляция, если изменения в линиях D~NaI произошли на день раньше, чем изменения в Н$\alpha$. Сдвигая ряд данных в другую сторону, проверяем обратную гипотезу. Таким образом мы определяем -- что является причиной, а что следствием. 

Next, we use the time series cross-correlation method. We create a uniform data series [HJD, D1r] with a step of one day, and the same data series [HJD, Fb]. Dates on which there were no observations contain empty cells in the series. We plot a graph of Fb as a function of D1r and determine the correlation coefficient. Then, we shift the series [HJD, D1r] by one day relative to the series [HJD, Fb], so that the Fb parameter is compared with the D1r parameter on the previous day, plot the graph again, and determine the correlation coefficient. In this way, we check: will the correlation changes if chages in the D~NaI lines occur one day earlier than changes in Н$\alpha$. By shifting the data series in the other direction, we test the inverse hypothesis. This way, we determine what is the cause and what is the effect.

%\textbf{Поскольку характерное время изменений в потоках аккреции и ветра составляет несколько суток, при любом сдвиге наблюдается положительная корреляция: чем сильнее поток аккреции, тем сильнее поток ветра, но дисперсия точек на этой зависимости довольно большая. Методом кросс-корреляции временных рядов мы пытаемся определить, при каком сдвиге наблюдается максимальная корреляция.} 

Since the characteristic timescale for changes in accretion and wind fluxes is several days, a positive correlation is observed at any shift: the stronger the accretion flux, the stronger the wind flux. However, the dispersion of data points on this curve is quite large. Using time series cross-correlation, we attempt to determine at what shift the maximum correlation is observed.

%\textbf{Наши наблюдения проводились короткими сетами по 5-7 ночей подряд в течение каждого месяца, с сентября по март. Исследовать корреляцию  в таких фрагментированных рядах данных можно лишь при сдвигах временных рядов в пределах  не более $\pm$4 суток. При этом, при нулевом сдвиге в корреляции участвуют все ночи наблюдений, но при сдвиге на 3-4 суток число данных значительно сокращается, и из-за этого возрастает ошибка определения коэффициента корреляции.}

Our observations were conducted in short sets of 5-7 consecutive nights during each month, from September to March. Correlation in such fragmented data series can only be investigated with time series shifts of no more than $\pm$4~days. At zero shift, all observation nights are included in the correlation, but with a shift of 3-4 days, the number of data points is significantly reduced, increasing the error in determining the correlation coefficient.

%\textbf{На Рис. 5 показано как меняется коэффициент корреляции в зависимости от сдвига временных рядов аккреции и ветра относительно друг друга.
%Характерный изгиб этой зависимости (максимум при D1r = -2d и минимум при D1r = +2d) указывает на то, что временной ряд аккреции смещен на два дня относительно  временного ряда ветра -- события аккреции опережают события ветра. Это дает основание предположить, что аккреция является причиной, а ветер -- следствием.}

Fig.5 shows how the correlation coefficient varies depending on the relative shift of the accretion and wind time series. The characteristic bend in this dependence (maximum at D1r =--2d and minimum at D1r = +2d) indicates that the accretion time series is shifted by two days relative to the wind time series, i.e. accretion events lead wind events. This suggests that accretion is the cause and wind is the effect.

\begin{figure}
\includegraphics[scale=0.8]{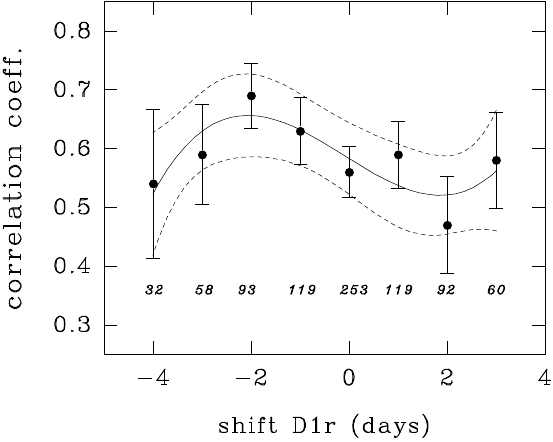}
\caption{Cross-correlation of the [HJD, D1r] and [HJD, Fb] time series with a shift of one series relative to the other within the range of -4 to +3 days. The solid line is the approximation by a third-degree polynomial. The dotted lines show the 95\% confidence interval. The maximum correlation corresponds to a shift of -2 days: first, the accretion changes, then the wind changes.}
\label{fig:a5}
\end{figure}

%\textbf{Бары ошибок, указанные на рисунке, определены формально как $\sqrt{(1-r^2)/(n-2)}$, где r - коэффициент корреляции, n - число наблюдений.}

The error bars shown in the figure are formally defined as $\sqrt{(1-r^2)/(n-2)}$, where r = correlation coefficient, n = number of measurements.

%На Рис. 6 показана зависимость между параметрами аккреции и ветра при разных сдвигах временного ряда [HJD, D1r] относительно временного ряда  [HJD, Fb].  При сдвиге на 2 дня число данных, участвующих в корреляции,  значительно сокращается,  но линия регрессии остается в пределах доверительного интервала >99\%. 

Fig.6 shows the relationship between accretion and wind parameters for different shifts of the [HJD, D1r] time series relative to the [HJD, Fb] time series. With a shift of 2 days, the number of data points involved in the correlation is significantly reduced, but the regression line remains within the confidence interval >99\%.

\begin{figure*}
\includegraphics[scale=0.63]{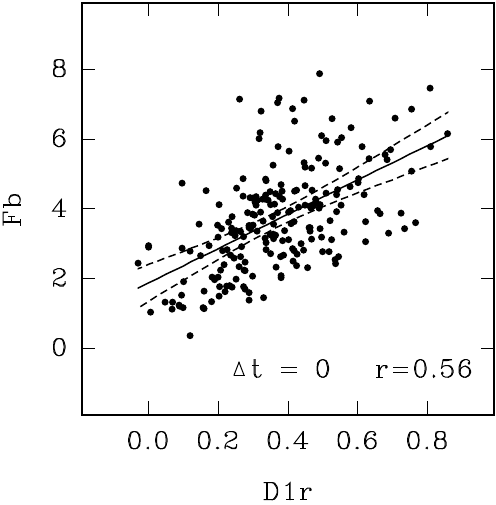}
\includegraphics[scale=0.63]{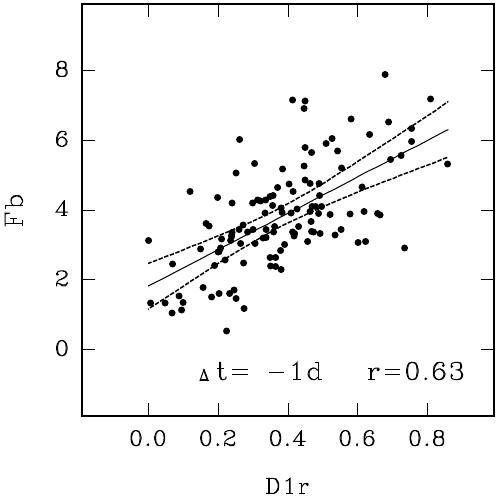}
\includegraphics[scale=0.63]{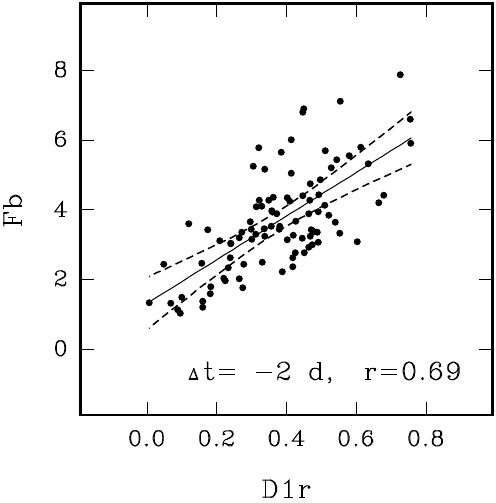}
\caption{Correlation between the accretion (D1r) and wind (Fb) parameters at different time series shifts: the maximum correlation coefficient (r = 0.69) was obtained at $\Delta$t =-2~days: accretion changes first, and then the wind changes 2~days later. The dotted lines indicate the 99\% confidence interval.}
\label{fig:a6}
\end{figure*}

%Можно было бы предположить, что сдвиг временных рядов аккреции и ветра вызван осевым вращением звезды с аксиально несимметричной магнитосферой: с одной стороны мы наблюдаем преобладающую аккрецию,  а через пол-оборота звезды  видим преобладающий ветер. Но в этом случае мы регистрировали бы периодические вариации этих спектральных признаков с периодом вращения звезды. Однако, такой периодичности не наблюдается. Частотный анализ показывает, что в изменениях параметров аккреции (D1r) и ветра  (Fb) нет периода вращения звезды. И, как указывалось выше, в интервалах непрерывных наблюдений в течение 6-7 суток (два оборота звезды)  не было существенных изменений профилей линий.

One might assume that this is an effect of the axial rotation of a star with an axially asymmetric magnetosphere: on one side, we observe dominant accretion, and after half a rotation of the star, we see dominant wind. However, in this case, we would register periodic variations in these spectral signatures with the rotation period of the star. However, no such periodicity is observed. Frequency analysis shows that the variations in the accretion (D1r) and wind (Fb) parameters do not correspond to the rotation period of the star. And, as noted above, during continuous observation intervals of 6-7 days (two rotations of the star), there were no significant changes in the line profiles.

\section{Discussion}
%Корреляции между событиями аккреции и ветра CTTS обнаруживаются в программах спектрального мониторинга. Наиболее детальное исследование такого рода было проведено при наблюдении AA~Tau \citep{Bou2003}.% (Bouvier et al. 2003). 
%~Было показано, что деформация силовых линий магнитосферы звезды вследствие дифференциального вращения магнитосферы и аккреционного диска приводит к повторяющимся событиям перестройки и восстановления магнитосферы. Это отражается в динамике газовых потоков аккреции и ветра и проявляется в переменности профилей  эмиссионных линий в спектре звезды \citep{Bou2003}.% (Bouvier et al. 2003). 
%~В случае AA~Tau xарактерная временная шкала таких изменений составляет около месяца. 

Correlations between CTTS accretion and wind events are detected in spectral monitoring programs. The most detailed study of this kind was conducted during observations of AA~Tau \citep{Bou2003}.% (Bouvier et al. 2003). 
~It was shown that the deformation of the star's magnetospheric field lines due to differential rotation of the magnetosphere and accretion disk leads to repeated magnetospheric reorganization and recovery events. This is reflected in the dynamics of the accretion and wind gas flows and manifests itself in variability in the emission line profiles in the star's spectrum \citep{Bou2003}.%(Bouvier et al. 2003). 
~In the case of AA~Tau, the characteristic timescale of such changes is approximately a month.

%В отличие от AA~Tau,  RY~Tau является более массивной ($\sim$2M$_\odot$), быстро вращающейся (P\approx3$^d$) звездой. Радиус магнитосферы близок к радиусу коротации и, возможно, превышает его. В этом случае возникает режим магнитного пропеллера и конический ветер \citep{Rom2009}.%(Romanova et al. 2009). 
%~Такой ветер стартует с границы магнитосферы и аккреционного диска  и ускоряется  силой градиента магнитного давления тороидальных полей \citep{Rom2009, Takasao2022}.% (Romanova et al. 2009; 2013;  Takasao et al. 2022). 
%~По мере удаления от звезды, конический ветер коллимируется в джет.

Unlike AA~Tau, RY~Tau is a more massive ($\sim$2M$_\odot$) and rapidly rotating (P$\approx3^d$) star. The magnetospheric radius is close to the corotation radius and possibly exceeds it. In this case, a magnetic propeller regime and a conical wind arise \citep{Rom2009}.%(Romanova et al. 2009). 
~The conical wind originates from the magnetosphere-accretion disk boundary and is accelerated by the magnetic pressure gradient of the toroidal fields \citep{Rom2009, Takasao2022}.% (Romanova et al. 2009; 2013; Takasao et al. 2022). 
~As it moves away from the star, the conical wind collimates into a jet.

%Наблюдаемая корреляция между изменениями потоков аккреции и ветра RY~Tau означает, что максимальная плотность ветра находится вблизи границы магнитосферы. Это предполагает вариант \textit{конического} ветра. Дисковый ветер, стартующий с более протяженной поверхности аккреционного диска, не мог бы обеспечить наблюдаемую корреляцию между событиями аккреции и ветра на шкале времени в два дня. Мы смотрим на звезду свозь оболочку конического ветра,  видим дискретные события аккреции газа на звезду и довольно быструю реакцию ветра на эти события.  

The observed correlation between the variations in accretion and wind fluxes in RY~Tau indicates that the maximum wind density is located near the magnetospheric boundary. This supports the \textit{conical} wind scenario. A disk wind originating from the more extended surface of the accretion disk could not provide the observed correlation between accretion and wind events on a two-day timescale. We are looking at the star through the conical wind envelope, seeing discrete events of gas accretion onto the star and the relatively rapid response of the wind to these events.

%В случае RY~Tau, наблюдаемое запаздывание реакции ветра на изменения аккреции может быть объяснено геометрией ветра. В момент усиления аккреции прекращается истечение ветра на границе магнитосферы. Глядя на звезду под углом 60-65\,$^{\circ}$ к оси вращения диска, мы увидим падение газа в проекцию на звезду уже в течение суток, но изменение плотности ветра на луче зрения произойдет позже, когда ветер, стартовавший с границы магнитосферы, поднимется над поверхностью диска и окажется на луче зрения в проекции на звезду (см. Рис. 7).  При средней скорости $\sim$100~км\,с$^{-1}$ ветер за два дня пройдет расстояние  $\sim$0.11~а.е., что равно примерно двум радиусам магнитосферы RY~Tau. 

In the case of RY~Tau, the observed delay in the wind's response to accretion changes can be explained by the wind's geometry. At the moment of accretion intensification, wind outflow at the magnetosphere's boundary ceases. Looking at the star at an angle of 60-65\,$^{\circ}$ to the disk's rotation axis, we will see gas infall projected onto the star within a day, but the change in wind density along the line of sight will occur later, when the wind, originating from the magnetosphere's boundary, rises above the disk's surface and appears along the line of sight projected onto the star (see Fig. 7). At an average velocity of $\sim$100~km\,s$^{-1}$, the wind will travel a distance of $\sim$0.11~AU in two days, which is approximately twice the radii of RY~Tau's magnetosphere.

\begin{figure}
\includegraphics[scale=1]{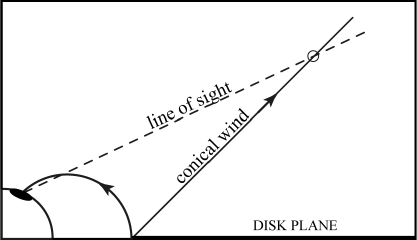}
\caption{Star, magnetosphere, and the conical wind vector.}
\label{fig:a7}
\end{figure}

%Особый интерес вызывает тот факт, что преобладающее направление потока газа на луче зрения (аккреция или ветер) меняется с характерным временем около 20~суток.  За время одного сета наблюдений (5-7~суток) направление потока как правило не меняется, но за время между сетами (20-30~суток) почти всегда меняется. 

Of particular interest is the fact that the dominant direction of gas flow along the line of sight (accretion or wind) varies with a characteristic time of approximately 20~days. During a single observation set (5-7~days), the flow direction generally remains constant, but during the time between sets (20-30~days), it almost always changes.

%Это похоже на работу “неустойчивого пропеллера”, когда радиус магнитосферы не постоянен. Радиус магнитосферы определяется балансом давления магнитного поля звезды ($\sim$B$^{2}/8\pi$) и  динамического давления газового потока ($\sim$$\rho$v$^{2}/2$) в аккреционном диске. Магнитное поле звезды (B) не меняется столь быстро, но в плотности газа ($\rho$) на внутренней границе аккреционного диска могут происходить изменения, поскольку диск не является однородным -- присутствие формирующихся планет  в диске вызывает волны плотности, которые достигают внутренней границы диска (см, например, Armitage 2010). %Armitage, 2010).  
%~Изменение плотности влияет на положение границы магнитосферы и, как следствие, может изменить режим магнитного пропеллера. 

This is similar to the operation of an “unstable propeller”, where the radius of the magnetosphere is not constant. The radius of the magnetosphere is determined by the balance between the pressure of the star's magnetic field ($\sim$B$^{2}/8\pi$) and the dynamic pressure of the gas flow ($\sim$$\rho$v$^{2}/2$) in the accretion disk. The star's magnetic field (B) does not change as rapidly, but the gas density ($\rho$) at the inner boundary of the accretion disk can change because the disk is not uniform -- the presence of forming planets in the disk causes density waves that reach the inner boundary of the disk (see, e.g., Armitage, 2010).
~Density changes affect the position of the magnetospheric boundary and, as a result, can alter the magnetic propeller regime.

%Рассматривался также вариант односторонней аккреции, когда падение газа на звезду идет с одной стороны от  плоскости диска,  а ветер (и джет) стартуют с другой стороны, причем эти направления спонтанно переключаются с характерным временем около 30~суток (результат 2D-моделирования, см.  \citealt{Lovelace2010}).% Lovelace et al., 2010).  
%~Если  в аккреционном диске есть планета на эксцентричной или наклонной орбите, то она может синхронизировать этот процесс с периодом орбитального движения.  В случае RY~Tau,  период 21.6~суток был замечен в изменениях плотности ветра на луче зрения по наблюдениям 2013-2020~гг. (\cite{PetrovRo2021},% [Petrov et al. 2021] 
%~см. также \citealt{Vedula2024}).% Vedula  Johns-Krull,  2024).  
%~Кеплеровский период P=21.6$^{d}$ соответствует расстоянию  0.2~а.е., что равно четырем радиусам магнитосферы RY~Tau. В общем случае,  если планета не одна, то интерференция волн в аккреционном диске приведет к более сложной картине переменности.

The one-sided accretion scenario was also considered, where gas infalls onto the star from one side of the disk plane, while the wind (and jet) originates from the other side. These directions spontaneously switch with a characteristic time of approximately 30~days (result of 2D modeling, \citealt{Lovelace2010}).% Lovelace et al., 2010). 
~If the accretion disk contains a planet on an eccentric or inclined orbit, it can synchronize this process with the orbital period. In the case of RY Tau, a period of 21.6 days was observed in line-of-sight wind density variations based on observations from 2013 to 2020 (\citealt{PetrovRo2021},%Petrov et al. 2021] 
~see also \citealt{Vedula2024}).%Vedula & Johns-Krull, 2024).
~A Kepler period of P=21.6$^{d}$ corresponds to a distance of 0.2~AU, which is equal to four magnetospheric radii of RY~Tau. In general, if there is more than one planet, wave interference in the accretion disk will lead to a more complex variability pattern.

%\begin{acknowledgments}
\section{acknowledgments}
%Авторы выражают благодарность за оценки блеска RY~Tau из международной базы данных AAVSO, предоставленные наблюдателями со всего мира и использованные в этом исследовании.  
%Авторы выражают благодарность С.Ю.~Горда, A.A.~Djupvik, J.F.~Gameiro и D.E.~Mkrtichian за участие в программах спектральных наблюдений RY~Tau. 
%Авторы благодарны M.M.~Романовой за обсуждение статьи и ценные замечания. 
%\textbf{Авторы выражают благодарность анонимным рецензентам за внимательное прочтение статьи и замечания, которые позволили улучшить представление результатов.}

The authors thank  S. Yu. Gorda for the spectra obtained at the Kourovka Observatory, A.~A. Djupvik for the spectra obtained with the NOT telescope, J.~F. Gameiro for the spectra obtained with the CAHA telescope, and D.~E. Mkrtichian for the spectra obtained with the TNT telescope.

The authors are grateful for the brightness estimates of RY~Tau from the AAVSO international database, provided by observers from around the world and used in this study.

The authors are grateful to M.~M. Romanova for discussion of the article and valuable comments.

%\end{acknowledgments}

\bibliographystyle{aspb1}
\bibliography{timeLag.bib}

%Верстка английского абстракта статьи:
\onecolumngrid
 \clearpage

\selectlanguage{english}
\begin{center}\bfseries Time lag between accretion and wind events in the T Tauri star RY~Tau
\end{center}
\begin{center}
\bfseries E.~V.~Babina$^1$, P.~P.~Petrov$^1$, K.~N.~Grankin$^1$, S.~A.~Artemenko$^1$
\end{center}
\begin{center}
%\bfseries
 %\small
 \footnotesize
 %\scriptsize
 %\tiny

 $^1$Crimean Astrophysical Observatoty, RAS,  Repulic of Crimea, 298409 Russia\\
\end{center}
\begin{center}
\begin{minipage}{\textwidth - 2cm}
\small Results of spectroscopic and photometric monitoring of the Classical T Tauri type star RY~Tau are presented. The observations cover a time interval since 2013 to 2024, a total of 220 nights. During the observations, the brightness of the star varied within the range of  V=9-11$^{m}$. The rotation axis of the "star + accretion disk" system is inclined at a large angle, so that the line of sight intersects both the region of the wind and the accreting flows  in the magnetosphere of the star. 
Variability of the flux in the shortwave wing of the H$\alpha$ emission line  and variability of the D\,NaI resonance doublet profile are analyzed.  It is shown that the wind and accretion fluxes vary on a time scale of about 20-24~days.  When the flow direction changes, a time lag is observed: first, accretion increases,  and after two days, absorption in the wind along the line of sight decreases. 
It is concluded that the spectral line profiles are formed in magnetospheric accretion flows  and in the conical wind starting from the magnetosphere boundary.  The time lag is determined by 
the tilt of the magnetic dipole and the opening angle of the conical wind. It is suggested that RY~Tau is in an unstable propeller mode, with fluctuations in the accretion and wind flows caused by density waves in the accretion disk.
\end{minipage}
\end{center}

\begin{center}
\begin{minipage}{\textwidth - 2cm}

Keywords: {\it Stars: variables: T Tauri, Herbig Ae/Be – Stars: winds, outflows – Line: profiles – Stars: individuals: RY~Tau}
\end{minipage}
\end{center}
\selectlanguage{english}

\end{document}

%% file: sao_cmd_author.tex
\def\squareforqed{\hbox{\rlap{$\sqcap$}$\sqcup$}}

\def\sq{\ifmmode\squareforqed\else{\unskip\nobreak\hfil
\penalty50\hskip1em\null\nobreak\hfil\squareforqed
\parfillskip=0pt\finalhyphendemerits=0\endgraf}\fi}

\def\utw{\smash{\rlap{\lower5pt\hbox{$\sim$}}}}

\def\udtw{\smash{\rlap{\lower6pt\hbox{$\approx$}}}}

\def\diameter{{\ifmmode\mathchoice
{\ooalign{\hfil\hbox{$\displaystyle/$}\hfil\crcr
{\hbox{$\displaystyle\mathchar"20D$}}}}
{\ooalign{\hfil\hbox{$\textstyle/$}\hfil\crcr
{\hbox{$\textstyle\mathchar"20D$}}}}
{\ooalign{\hfil\hbox{$\scriptstyle/$}\hfil\crcr
{\hbox{$\scriptstyle\mathchar"20D$}}}}
{\ooalign{\hfil\hbox{$\scriptscriptstyle/$}\hfil\crcr
{\hbox{$\scriptscriptstyle\mathchar"20D$}}}}
\else{\ooalign{\hfil/\hfil\crcr\mathhexbox20D}}%
\fi}}